\documentclass[journal]{IEEEtran}
\usepackage{amsmath,amssymb,amsfonts}
\usepackage{pifont}

\usepackage{algorithmic}
\usepackage[linesnumbered,ruled,vlined]{algorithm2e}
\usepackage{amsthm}
\usepackage{graphicx}
\usepackage{textcomp}
\usepackage{paralist}
\usepackage{subcaption}
\usepackage{hyperref}
\usepackage{multirow}
\usepackage{url}
\usepackage{setspace}
\usepackage{tabu}
\usepackage{romannum}
\usepackage{tablefootnote}
\usepackage{booktabs}
\usepackage{enumitem}
\usepackage{arydshln}
\usepackage[normalem]{ulem}
\usepackage{threeparttable}
\usepackage{flushend}
\usepackage{url}
\usepackage{xcolor}
\usepackage{soul}

\newcolumntype{L}[1]{>{\raggedright\let\newline\\\arraybackslash\hspace{0pt}}m{#1}}
\newcolumntype{C}[1]{>{\centering\let\newline\\\arraybackslash\hspace{0pt}}m{#1}}
\newcolumntype{R}[1]{>{\raggedleft\let\newline\\\arraybackslash\hspace{0pt}}m{#1}}

\newcolumntype{?}{!{\vrule width 1pt}}

\def\BibTeX{{\rm B\kern-.05em{\sc i\kern-.025em b}\kern-.08em
    T\kern-.1667em\lower.7ex\hbox{E}\kern-.125emX}}

\makeatletter
\newcommand{\thickhline}{%
	\noalign {\ifnum 0=`}\fi \hrule height 1pt
	\futurelet \reserved@a \@xhline
}
\newcolumntype{"}{@{\hskip\tabcolsep\vrule width 1pt\hskip\tabcolsep}}
\makeatother

\begin{document}
%
\title{An Empirical Study of In-App Advertising Issues Based on Large Scale App Review Analysis}

\author{Cuiyun~Gao,~Jichuan~Zeng,~David~Lo,~Xin~Xia,~Irwin~King,~\IEEEmembership{Fellow,~IEEE}
        and~Michael~R.~Lyu,~\IEEEmembership{Fellow,~IEEE}
\thanks{C. Gao was with the School of Computer Science and Technology, Harbin Institute of Technology (Shenzhen), China (e-mail: gaocuiyun@hit.edu.cn)}
\thanks{J. Zeng, I. King, and M. R. Lyu were with the Department of Computer Science and Engineering, The Chinese University of Hong Kong, Hong Kong, China (email: \{jczeng, king, lyu\}@cse.cuhk.edu.hk)}
\thanks{D. Lo is with the School of Information Systems, Singapore Management University, Singapore. (email: davidlo@smu.edu.sg)}
\thanks{ X. Xia is with the Faculty of Information Technology, Monash University, Australia. (email: xin.xia@monash.edu)}
\thanks{J. Zeng is the corresponding author.}

\thanks{Manuscript received August 22, 2020.}}

%
%

\markboth{Journal of \LaTeX\ Class Files,~Vol.~14, No.~8, August~2015}%
{Shell \MakeLowercase{\textit{et al.}}: Bare Demo of IEEEtran.cls for IEEE Journals}
%



\maketitle

\begin{abstract}
In-app advertising closely relates to app revenue. Reckless ad integration could adversely impact app reliability and user experience, leading to loss of income. It is very challenging to balance the ad revenue and user experience for app developers.

In this paper, we present a large-scale analysis on ad-related user feedback. The large user feedback data from App Store and Google Play allow us to summarize ad-related app issues comprehensively and thus provide practical ad integration strategies for developers. We first define common ad issues by manually labeling a statistically representative sample of ad-related feedback, and then build an automatic classifier to categorize ad-related feedback.
We study the relations between different ad issues and user ratings to identify the ad issues poorly scored by users. We also explore the fix durations of ad issues across platforms for extracting insights into prioritizing ad issues for ad maintenance. 

We summarize 15 types of ad issues by manually annotating 903/36,309 ad-related user reviews. From a statistical analysis of 36,309 ad-related reviews, we find that users care most about the number of unique ads and ad display frequency during usage. Besides, users tend to give relatively lower ratings when they report the security and notification related issues. Regarding different platforms, we observe that the distributions of ad issues are significantly different between App Store and Google Play. Moreover, some ad issue types are addressed more quickly by developers than other ad issues. We believe the findings we discovered can benefit app developers towards balancing ad revenue and user experience while ensuring app reliability.
\end{abstract}


%
\IEEEpeerreviewmaketitle

\section{Introduction}\label{sec:intro}
In-app advertising is a type of advertisement (ad) within mobile applications (apps). Many organizations have successfully monetized their apps with ads and reaped huge profits. For example, the mobile ad revenue accounted for 76\% of Facebook's total sales in the first quarter of 2016~\cite{facebook}, and increased 49\% year on year to about \$10.14 billion in 2017~\cite{fbadincrease}. 
Triggered by such tangible profits, mobile advertising has experienced tremendous growth recently~\cite{adreport}. Many free apps, which occupy more than 68\% of the over two million apps in Google Play~\cite{market}, adopt in-app advertising for monetization. However, the adoption of ads has strong implications for both users and app developers. According to a survey in 2016~\cite{adsurvey}, almost 50\% of users uninstalled apps just because of ``intrusive'' mobile ads, resulting in a heavy reduction in user volume of the apps. Inappropriate ad integration could also increase the difficulty of ensuring app reliability~\cite{DBLP:conf/ccs/0001BD16,DBLP:journals/tr/ChenZCMLW20,DBLP:journals/tr/TaoZGL18}. Moreover, the reduced audiences would generate fewer impressions (\textit{i.e.}, display of ads) and clicks for in-app ads, thereby making developers harder to earn ad profits.


Past studies have conducted surveys to understand users' perceptions of mobile advertising, \textit{e.g.}, perceived interactivity~\cite{yu2013you}, usefulness~\cite{soroa2010factors}, and credibility~\cite{chowdhury2006consumer}. There also exists research devoted to investigating or mitigating the hidden costs of ads, \textit{e.g.}, energy~\cite{mohan2013prefetching}, traffic~\cite{nath2015madscope}, system design~\cite{grace2012unsafe}, and other factors~\cite{gui2015truth,sooel2016what}. Recent research resorts to user reviews to identify impact of in-app advertising on user experience and app reliability. For example, Ruiz \textit{et al.}~\cite{ruiz2014impact} analyze the impact of 28 ad libraries on ratings of Android apps. They find that integrating certain specific ad libraries can negatively affect app ratings. Gui \textit{et al.}~\cite{gui2015truth} also observe that ads can adversely impact user feedback, \textit{i.e.}, over 50\% of the studied ad-embedded apps have at least 3.28\% of their user complaints dealing with ads. However, few studies have been conducted to identify the common ad issues from app reviews in large scale.




In this paper, we conduct large-scale user review analysis for characterizing common ad issues and providing detailed insights into ad design and maintenance for developers while preserving app reliability. Specifically, we investigate 32 cross-platform apps that rank in the top 100 list of Apple's App Store\footnote{In this paper, App Store refers to Apple's App Store.} and Google Play, and examine the following 
research questions:

\begin{inparaenum}[\itshape RQ1:]
	\item \textit{What are the common types of ad issues in ad-related user feedback?} \par
	We answer the question by analyzing a large collection of ad-related user feedback. To determine the common ad issues, we first manually annotate a statistically representative feedback sample. We summarize 15 types of common ad issues based on the manual annotation. Then following a standard keyword-based approach~\cite{DBLP:conf/sigsoft/RayPFD14,DBLP:conf/wcre/KochharWL16} for automatic classification, we group the whole ad-related reviews into the 15 manually defined issue types. From the automatic analysis on 36,309 ad reviews, we find that users care most about the number of unique ads and ad display frequency among all the issue types.

	\item \textit{What are the relationships between ad issue types that users described in their ad-related reviews and the ratings that they gave?}\par 
	This question aims at helping developers focus on the ad issue types that users tend to be more negative about.
	In terms of absolute numbers, we discover that nearly half of the low-rated reviews, \textit{i.e.}, with star ratings in the range of one to three, talk about the number of ads and display frequency. Also, users are likely to give poor ratings to ad issues related to notification (\textit{e.g.}, ads notifying users through the status bar) and security (\textit{e.g.}, unauthorized data collection or permission usage) types, despite their lower percentages than other issue types.

	\item \textit{How different are the distributions of ad issue types in different platforms?}\par
	To expand the revenue and reach more users, app developers generally build cross-platform apps, making apps available on multiple platforms.
	By answering this question, developers can pay attention to the platform differences, and prioritize ad issue types according to platforms. We find that for each ad issue type, its distributions between App Store and Google Play are significantly different. 
	
	\item \textit{What types of ad issues are addressed more quickly by developers?}\par
	App developers would address the important app issues commented by users in the updated versions.
	Identifying the ad issue types prioritized by many app developers can give us insights for ad maintenance. We find that issue types related to the number of ads and ad contents are solved within relatively longer periods than other types. Moreover, we observe that comparing App Store and Google play, developers solve ad issues at significantly different speeds. Generally, ad issues reflected in Google Play are addressed more quickly than those in App Store.
\end{inparaenum}

Our study has implications for both developers and researchers. First, our study indicates the main ad issue types developers should pay attention to.
Our study also suggests that developers should pay attention to the platform difference during ad design and maintenance. Additionally, our study shows the existence of platform difference phenomenon (i.e., users respond differently when using the same app in different platforms), and this suggests an interesting direction of future work in platform-aware app design, testing and analysis (e.g., how to automatically customize and test apps for different platforms to improve app reliability).
The key contributions of our work are as follows.

\begin{inparaenum}[1\upshape)]
    \item We investigate common types of ad issues by analyzing a large user feedback corpus. \par
    \item We conduct statistical analysis on ad issue types by considering multiple factors, including user ratings, platforms, and the fix durations (\textit{i.e.}, the amount of time that has elapsed before the issue is fixed).\par
	\item We summarize the implications on better designing and maintaining ads for app developers.\par
	
\end{inparaenum}


\textbf{Paper structure.} Section~\ref{sec:method} presents the methodology we used for cost measurement and user review analysis. Section~\ref{sec:exper} describes the findings of our study. Section~\ref{sec:lessons} illustrates lessons we learned from review analysis for ad design and maintenance. Section~\ref{sec:dis} presents threats to validity. Related work
and final remarks are discussed in Section~\ref{src:literature}
and Section~\ref{sec:conclusion}, respectively.
\section{Study Procedure}\label{sec:method}
In this section, we elaborate on the study procedures we adopted for data collection and categorization of ad issue type..

\subsection{Data Collection}
We manually select 32 popular apps published on both App Store and Google Play from their respective top 100 free app list\footnote{We referred to the top charts provided by App Annie~\cite{appannie}.}. The apps are listed in Table~\ref{tab:apps}. The major consideration for the selection is the number of user feedback, \textit{i.e.}, the apps should have more than 100,000 reviews on both app stores. It can also be seen that the subject apps cover a broad range of categories (15 categories in total). After determining the apps, we built a simple web crawler to automatically collect the user reviews of these apps online.

In total, we downloaded 1,840,349 and 3,243,450 of user reviews for the 32 apps from App Store and Google Play, respectively (see Table~\ref{tab:dataset}). The reviews on both platforms were collected during September, 2014 to March, 2019. The discrepancy between the number of user reviews across the app stores is due to the different number of users and exposed data for collection on the platforms~\cite{DBLP:journals/ese/McIlroyAKH16}. We define ad-related reviews (ad reviews, for short) as those containing keywords related to ads, \textit{i.e.}, regex = ad/ads/advert*. In total, we identify 36,309 ad-related reviews. Although such ad review extraction method is not perfect, we hypothesize that the strong selection criterion can reduce false positives.


\begin{table}[t]
	\center
	\caption{Cross-platform subject apps.}
	\label{tab:apps}
	\scalebox{0.82}{\begin{tabular}{l |l?l| l}
	\hline
	\textbf{App Category} & \textbf{App Name} & \textbf{App Category} & \textbf{App Name}\\
	\hline
	\hline
	\multirow{3}{*}{Casual} & Candy Crush & Photography & Camera360\\
    \cline{3-4}
	& Minion Rush & \multirow{2}{*}{Education} & Duolingo\\
	& My Talking Tom & & TED\\
	\hline
	\multirow{2}{*}{Shopping} & eBay & Tools & SHAREit \\
	\cline{2-4}
	& Amazon & Music & SoundCloud \\
	\hline
	\multirow{4}{*}{Entertainment} & Netflix & Arcade & Subway Surfers \\
	\cline{2-4}
	& YouTube & Travel & TripAdvisor \\
	\cline{2-4}
	& Spotify Music & Trivia & Trivia Crack \\
	\cline{2-4}
	& VLC & \multirow{6}{*}{Communication} & Line \\
	\cline{1-2} \cline{4-4}
	\multirow{6}{*}{Social} & Facebook & & Messenger \\
	\cline{2-2} \cline{4-4}
	& Twitter & & Skype \\
	\cline{2-2} \cline{4-4}
	& Pinterest& & WeChat \\
	\cline{2-2} \cline{4-4}
	& Snapchat&  & WhatsApp \\
	\cline{2-2} \cline{4-4}
	& Tango & & Viber \\
	\cline{2-4}
	& Instagram & Transportation & HERE \\
	\hline
	Maps & Waze & Productivity & Evernote \\
	\hline
	\end{tabular}
	}
\end{table}	

\subsection{Categorizing Ad Issues}\label{subsec:classapproach}
We first introduce the manual process where we define ad issues based on a statistically representative sample of ad reviews, and then present the automated classification method we adopt for automatically classifying the whole ad reviews.


\subsubsection{Manual Categorization}\label{subsec:manualcate}
Users often leave important pieces of information in the feedback while complaining about ads. Such information may relate to the displaying style of ads, and in what way ads affect the functionalities of an app. To determine the ad complaint topics, we perform card sort~\cite{spencer2009card}. 

Card sorting is a technique that is widely adopted to derive taxonomies from data. We use card sorting here to summarize common ad issues that users complained about. Following the three phases of card sorting~\cite{DBLP:conf/icse/BegelZ14,DBLP:conf/icse/Kim0DB16}: In the \textit{preparation} phase, we 
select 903/36,309 ad reviews to give us a 95\% confidence level with 1\% confidence interval; in the \textit{execution} phase, reviews are sorted into meaningful categories with a descriptive textual label; finally, in the \textit{analysis} phase, hierarchies are formed in order to deduce general categories.
Specifically, our card sort was open, and we let the groups emerge and evolve during the sorting process. Similar to~\cite{DBLP:conf/icse/BegelZ14,DBLP:conf/icse/Kim0DB16}, the card sorting process was conducted by the first author and second author separately. Both card sorts led to similar categories of ad issues, with agreement rate at 97.1\%, and were finalized based on their joint discussion. Ultimately, this resulted in 15 ad issue types shown in Table~\ref{tab:category}. The ad issue types are further categorized into two large groups based on whether they are related to the ads (\textbf{In-Ad}) or the impact of ads on apps (\textbf{App}).

Grouping an ad review into the ``Other/Unknown'' type is usually based on the following reasons:
\textit{a)} Although the review contains the ad-related keywords using regex, it actually does not talk about in-app ads, e.g., the first two pieces of reviews in Table~\ref{tab:other};
\textit{b)} The in-app ad does not impact user's experience actually. For example, for the third and fourth reviews in Table~\ref{tab:other}, they state that he/she likes the free music even with ads, or the ad loads fine, respectively; \textit{c)} The review does not clearly state what aspect of the advertisement he/she does not like, or the review does it in a vague way, e.g., the last review in Table~\ref{tab:other} describes that the ad is ``annoying'' but does not describe in what way the ad is annoying. During manual analysis, 39.87\% (360/903) ad reviews are labeled as ``Other/Unknown'' type, which indicates a large proportion of \textit{non-useful} ad reviews.

\begin{table}[t]
	\center
	\caption{Statistics for data collected from App Store and Google Play.}
	\label{tab:dataset}
	\scalebox{1.0}{\begin{tabular}{|c | c|c | c |}
		\hline
		 \multicolumn{2}{|c|}{\textbf{All Reviews}} & \multicolumn{2}{c|}{\textbf{Ad Reviews}}  \\
		 \hline
		 \hline
		 App Store & Google Play & App Store & Google Play  \\
		 
		 \hline
		 1,840,349 & 3,243,450 & 22,343 & 13,966  \\
		\hline
	\end{tabular}
}
\end{table}

\subsubsection{Automated Classification}\label{subsec:auto_class}
Each ad review can be categorized into one or more than one issue type. For example, one ad review of a video player app, ``\textit{30 second adverts are not skippable and they can not be loaded properly leading to buffer... So advert is 2 to 3 minutes long}'', was complaining about the non-skippable and timing aspects of the video ads, and also the slow app functionality caused by the ads. We automatically categorize the ad issues of each ad review, taking a similar approach in Ray \textit{et al.}~\cite{DBLP:conf/sigsoft/RayPFD14}.
This automated classification is performed in two phases: Keyword matching and supervised multi-label classification.

\begin{table}[t]
	\center
	\caption{Ad review examples that are labeled as ``Other/Unknown'' Type.}
	\label{tab:other}
	\scalebox{0.9}{\begin{tabular}{|L{1.2cm}|L{1.6cm}|L{3.7cm}|c|}
		\hline
		\textbf{App Name} & \textbf{Title} & \textbf{Review Text} & \textbf{Star} \\
		\hline
		Spotify Music & As \underline{advertised} & I've only used it when hooked to wifi, but so far this app has been awesome. & 5 \\
		\hline
		Netflix & Now works & Since the last update, I had problems starting the app, remained to load the splash screen \underline{ad infinitum}\tablefootnote{Ad infinitum is a Latin phrase meaning ``to infinity'' or ``forevermore''~\cite{adinfinitum}.}, then I deleted and reinstalled ...& 5 \\
		\hline
		Spotify Music & Love Spotify & ... Feels like free music even if I don't have the \underline{free ads version}. & 5 \\
		\hline
		YouTube & Unusable & It took me 1 hour to watch a 10 minute video because it either stops loading or stops playing all together. Whats worse is any other time I try and watch a video it doesn't even load. But \underline{the ads load fine} & 1 \\
		\hline
		Spotify Music & Emt & The free version has \underline{annoying ads} and limitations, but certainly a good premium. & 4
		 \\\hline
	\end{tabular}
}
\end{table}

\noindent \textbf{Step 1: Keyword matching.} 
We first use a keyword-based search technique to automatically categorize the ad reviews with potential ad complaint types. Since the ultra imbalanced distribution of categories, which might introduce too much bias for training multi-label classifier~\cite{ali2015classification}, we sample up to 280 training instances for each categories. After removing the duplicated multi-labeled reviews, we have 3,630 ad reviews as our training data. The ad reviews vary in length, from several words to hundreds of words. Since a large proportion of review texts may cover a wider range of app issues besides the ad-related ones, we focus on the ad-related sentences, \textit{i.e.}, the sentences containing keywords related to ads (regex=ad/ads/advert*), instead of the whole reviews. 

We choose a restrictive set of keywords and phrases as shown in Table~\ref{tab:category}. For example, if the ad-related sentences 
contain any of the keywords: loud, screech, play sound, or volume, we infer the review is related to the \textit{Volume} issue type. Such a restrictive set of keywords and phrases help to reduce false positives. 

\noindent \textbf{Step 2: Supervised multi-label classification.} 
We use the automatically annotated ad reviews from the previous step as training data for supervised learning of multi-label classification. We then use another manually annotated 
review set as our validation set for reporting the performance of our multi-label classifier.
We first tokenize each ad review into bag-of-words form, remove the common stop words provided by the NLTK toolkit\footnote{\url{https://www.nltk.org/nltk_data/}}. Then, we lemmatize each word of the review using the popular WordNet Lemmatizer and convert each ad review into tf-idf feature vector. Finally, we utilize Classifier Chains (CC)~\cite{DBLP:journals/ml/ReadPHF11} approach to transform the problem of classifying multi-labeled data into one or more problems of single labeling, and use the well-known Support Vector Machine (SVM)\footnote{\url{https://scikit-learn.org/stable/modules/svm.html}} as the basic estimator of CC to build a classifier based on the training data and to classify the remaining ad reviews.
\section{Findings}\label{sec:exper}
In this section, we try to answer the research questions illustrated in Section~\ref{sec:intro} and elaborate on our findings.

\subsection{RQ1: What are the common types of ad issues?}\label{subsec:rq1}
\subsubsection{Motivation}
Users play an essential role in the ad-profiting process, since the number of ads viewed or clicked by users determines the ad revenue: User retention and user base are critical for app developers. However, embedding ads inappropriately can ruin user experience. According to a survey in 2016~\cite{adannoying}, two in three app users consider mobile ads annoying and tend to uninstall those apps or score them lower to convey their bad experience. Such negative feedback is likely to influence other potential users, which further leads to customer churn and reduced ad revenue. This motivates us to capture the complained ad issues
in ad reviews, and draw developers' attention to the problematic aspects of ad usage.


\subsubsection{Method}
We first evaluate the multi-labeling classifier introduced in Section~\ref{subsec:classapproach}, and then use the trained classifier to automatically annotate the whole ad review corpus.
For evaluating the classifier, we manually labeled another 280 ad reviews. We compare the result of the automatic classifier with the manual annotation using the precision and recall as evaluation metrics~\cite{DBLP:journals/ese/McIlroyAKH16}. Precision for an issue type refers to the proportion of ad reviews that are correctly assigned to the type, among those that are assigned to the type. Recall for an issue type refers to the proportion of ad reviews that are correctly assigned to the type, among those that actually belonging to the type. 


\subsubsection{Findings}
Table~\ref{tab:result} summarizes the result for each ad issue type. We observe that the precision and recall are acceptable (more than 80\%). We then use the built classifier to categorize all the ad reviews. Table~\ref{tab:category} summarize the total numbers and percentages of ad reviews classified into each issue type.
We remove the 18,007 reviews grouped into ``Other/Unknown'' category, which leaves us with 18,302/36,309 reviews.

\textbf{Users complain most about the number of ads and ad display frequency.} From the reviews clearly expressing ad issues, we observe that most of the ad reviews complain about the number of ads (45.51\%) and display frequency (25.02\%). Although in-app advertising is an effective monetization strategy for mobile developers, too many ads and their frequent display can severely degrade user experience. For example, one user complained that ``\textit{There are too many ads whenever I try to switch to a different song after an ad. Make it stop. it's really annoying}''. Some apps provide reward ads, i.e., offer something to the user in exchange for watching or interacting with an ad. One example is Spotify Music. The users can enjoy 30-minute ad free music streaming by watching an ad video in the app. Such reward ads would be less unfavorable to users. For instance, one user commented that ``\textit{...You can still listen to music if you're ok with ads every once in awhile. 30 seconds worth of ads, it isn't that bad}''. According to one survey in 2018~\cite{rewardedads}, reward ads were rated as the most effective for delivering the best user experience by the majority of survey respondents. \textbf{Thus, developers could design a reward strategy to alleviate users' dislike for ads.} Additionally, the ads' display frequency should be set at a comfortable rate for app usage. For instance, one YouTube user stated that ``\textit{There are too many ads in a video that is 30 minutes and there are 7 ads}'', and gave one-star rating. \textbf{Developers could devise an A/B testing experiment to determine an optimal frequency ads should be displayed in an app.} 

\begin{table}[t]
	\center
	\caption{Multi-label classifier precision and recall results.}
	\label{tab:result}
	\scalebox{0.85}{\begin{tabular}{|l|l|r r|r r|}
		\hline
		 & \textbf{Ad Issue Type} & \begin{tabular}[x]{@{}c@{}}\textbf{\#Training}\\ \textbf{Data}\end{tabular} & \begin{tabular}[x]{@{}c@{}}\textbf{\#Test}\\ \textbf{Data}\end{tabular} &\textbf{Precision} & \textbf{Recall} \\
		 \hline
		 \hline
		 \multirow{10}{*}{\textbf{In-Ad}} & \textbf{Content} & 280 & 16 & 91.45\% & 77.78\%\\
        & \textbf{Frequency} & 280 & 18 & 89.74\% & 85.37\%  \\ 
        & \textbf{Popup} & 280 & 21 & 92.44\% & 100.00\%\\ 
        & \textbf{Too Many} & 280 & 86 & 91.94\% & 93.44\%\\ 
        & \textbf{Non-skippable} & 280 & 14 & 100.00\% & 72.73\%\\ 
        & \textbf{Timing} & 280 & 21 & 82.71\% & 88.44\% \\
        & \textbf{Size} & 280 & 17 & 83.31\% & 100.00\% \\
        & \textbf{Position} & 280 & 12 & 79.36\% & 90.48\%  \\
        & \textbf{Auto Play} & 280 & 16 & 100.00\% & 100.00\%  \\
        & \textbf{Volume} & 119 & 10 & 100.00\% & 80.00\% \\
        \hline
        \hline
        \multirow{5}{*}{\textbf{App}} & \textbf{Security} & 280 & 11 & 81.58\% & 89.26\%  \\
        & \textbf{Crash} & 280 & 15 & 75.17\% & 94.73\% \\
        & \textbf{Slow} & 280 & 6 & 100.00\% & 83.33\% \\
        & \textbf{Notification}  & 280 & 3 & 100.00\% & 100.00\% \\
        & \textbf{Orientation}  & 98 & 2 & 100.00\% & 100.00\%\\
		\hline
		\hline
		\multicolumn{2}{|c|}{\textbf{Average}}  & 254.46 & 20.33 & 91.18\% & 85.57\% \\
		\hline
	\end{tabular}
}
\end{table}

\begin{table*}[t]
	\center
	\caption{Categories and distribution of ad issue
	types for the 18,302 reviews which are not grouped in the ``Other/Unknown'' category.
	}
	\label{tab:category}
	\scalebox{0.93}{\begin{tabular}{ll L{4cm} L{7.4cm} r r}
		\hline
         \multicolumn{2}{c}{\textbf{Ad Issue Type}} & \textbf{Ad Issue Description} & \textbf{Search Keywords/Phrases} & \textbf{Count} & \textbf{\%Count} \\
        \hline
        \hline
        \multirow{10}{*}[-6em]{\textbf{In-Ad}} & \textbf{Content} & What is in the ads shown to users & irrelevant, same ad, open install page, target advertisement, random ad, advertise what & 1,613 & 8.81\%\\ \addlinespace[0.1cm]
        & \textbf{Frequency} & How often ads appear in an app & every time, ad rate, continuously, occasional ad, more than once, constantly, every \textless digit\textgreater\,second & 4,580 & 25.02\% \\  \addlinespace[0.1cm]
        & \textbf{Popup} & The way that ads suddenly appear to users &  pop, interruption, the middle of, half way through, onslaught, pop up, during a video, popup, suddenly, keep get in my way, interrupt & 2,475 & 13.52\% \\ \addlinespace[0.1cm]
        & \textbf{Too Many\tnote{*}} & How many ads are displayed to users & more ad, increase number, a few ad, ton of advertise, abundance of advertise, fill with ad, more and more, only with advertise, full of ad, much ad, a pile of advertise, more advertise, much advertise, some ad, lot ad, many, lot of ad, block& 8,329 & 45.51\% \\ \addlinespace[0.1cm]
        & \textbf{Non-skippable\tnote{*}} & Ads cannot be skipped by users & cant skip, be able to skip, skippable, wont stop, skip available & 703 & 3.84\% \\ \addlinespace[0.1cm]
        & \textbf{Timing} & Time interval of ad displaying & long, permanent, much time, short, never end, brief & 2,216 & 12.11\% \\\addlinespace[0.1cm]
        & \textbf{Size} & How big of an ad &  tiny, space, huge, half of the screen, banner & 385 & 2.10\% \\\addlinespace[0.1cm]
        & \textbf{Position} & Where an ad is placed & UI, bottom, too high, at the top, front page, button, in browser page, below & 1,233 & 6.74\% \\\addlinespace[0.1cm]
        & \textbf{Auto Play} & The way that ads start without permission & auto play, automatically play, auto skip & 359 & 1.96\% \\\addlinespace[0.1cm]
        & \textbf{Volume\tnote{*}} & Sound level of video or audio ads & loud, screech, play sound, volume & 159 & 0.87\% \\
        \hline
        \multirow{5}{*}[-3em]{\textbf{App}} & \textbf{Security} & Unauthorized data collection or permission usage & collect information, scam, private, virus, 
    access your camera, listen through, monitor & 341 & 1.86\% \\\addlinespace[0.1cm]
        & \textbf{Crash} & Apps not working caused by ads & black screen, doesnt work, doesnt load, turn black, not respond, dont work, freeze, stall, crash & 1,846 & 10.09\% \\\addlinespace[0.1cm]
        & \textbf{Slow} & Slow app functionalities caused by ads & buffer, laggy, delay, forever to load, for age, slow, for ever to load, take minute to load, lag, try to load, take time to load & 614 & 3.35\% \\\addlinespace[0.1cm]
        & \textbf{Notification} & Ads notifying users through the status bar & push ad, notification & 338 & 1.85\% \\\addlinespace[0.1cm]
        & \textbf{Orientation\tnote{*}} & The orientation of app screen impacted by ads & portrait, horizontal screen, landscape & 105 & 0.57\% \\
		\hline
	\end{tabular}}
\end{table*}

\textbf{Developers should pay attention to popup ads, ad timing, and ad content.} For the in-ad issues, we observe that reviews related to popup (13.52\%), timing (12.11\%) and content (8.81\%) also occupy large proportions among the whole ad reviews. Popup ads can effectively grab the attention of customers, but can also interrupt their interaction with apps. Popping ads in a video is popular among publishers that offer video content within their app~\cite{adformat}, and usually display with skippable or close options. For example, one three-star-rating review described that ``\textit{...There are times when I'm about to watch a video and an advert pops which I can choose to skip...}''. It is worth noting that the popup ads appearing during a call or when music is playing, etc., can lead
to extremely unpleasant experience for users. One Tango user commented that ``\textit{... right in the middle of a call there was an ear splitting sound. And when I looked at the phone screen there was an ad...}''. \textbf{Developers should be careful on introducing popup ads, especially ads with audio, such that they do not substantially reduce user experience.}

Too long ad display period and uninteresting content can also interfere with apps' usage for users. A one-star-rating review stated that ``\textit{...It's bad enough that I have to sit through a 30-sec ad that I'm even not interested in...}''. Long ad display periods and uninteresting content could try users' patience, and may drive potential users away. \textbf{Developers should provide skip option for long ads or consider better personalization to only present long ads with contents highly likely to be of interest to users.}

\textbf{Developers should notice ads' impact on apps' functionalities.} For the app-level issues, we find that crash (10.09\%) and slow response (3.35\%) issues are non-trivial in number among the whole ad reviews. Ad modules may be poorly implemented and incompatible with app functionality, resulting in app breakdown or slowing performance. For example, one user complained that ``\textit{Utterly disappointed with the current ads situation. They mess up AirPlay big time. Try airplaying to your TV. The moment ad starts, everything freezes}''. In this case, users cannot use the app properly. \textbf{Thus, developers should carefully integrate and test the ad libraries before deployment.} Besides the crash and slow response issues, 1.86\% complain about security, 1.85\% are related to notification through status bar, and 0.57\% are about app orientation being affected by ads. Users are less likely to complain about issues of these types.


\begin{center}
\setlength{\fboxrule}{1pt}
	\noindent\fbox{
		\parbox{0.46\textwidth}{

			\textbf{Finding 1:} Users care most (70.53\%) about the number of ads and ad appearing frequency among the ad issues. Other ad issue types such as the design of popup ads, ad timing, ad content, and crash also occupy obvious proportions among the ad reviews.

			}}
\end{center}

\subsection{RQ2: What are relationships between ad issue types that users described in their ad-related reviews and the ratings that they gave?}
\subsubsection{Motivation}
In RQ1, we identified the ad issue types commonly expressed via user feedback, and analyzed their quantity distributions. Besides review text, each ad review comes with a rating provided by the user on App Store and Google Play. Since user ratings influence how app platforms display apps in response to a user search, and have a great impact on the number of app downloads~\cite{DBLP:conf/msr/HarmanJZ12}, understanding the users' rating behavior when they complain about ad issues is important. We aim at identifying the ad issue types that are more likely to impact user ratings in this question.

\subsubsection{Methods}
To answer RQ2, we first divide the ad reviews into three polarities, i.e., positive, neutral, and negative, according to the given ratings, and then compare the quantity distributions of different ad issue types for the three sentiment polarities.
We consider the reviews with lower ratings (e.g., one or two) as negative reviews, the ones with higher ratings (e.g., four or five) as positive instances, and the others as neutral reviews.

We determine whether user ratings are independent of ad issue types by using Pearson's Chi-Squared test~\cite{pearson1900x} (or Chi-Squared test for short) at $\mathit{p-value}=0.05$~\cite{mchugh2013chi}.

We use Mann-Whitney U test~\cite{mann1947test}, a non-parametric test, to observe whether two issue types have significantly different rating distributions. We set the confidence level at 0.05 and apply the standard Bonferroni correction (which is the most conservatively cautious of all corrections) to account for multiple statistical hypothesis testing. To show the effect size of the difference between the two types, we compute Cliff's Delta (or d), which is a non-parametric effect size measure~\cite{DBLP:journals/technometrics/Ahmed06}. Following the guidelines in~\cite{DBLP:journals/technometrics/Ahmed06}, we interpret the effect size values as small for $0.147<d<0.33$, medium for $0.33<d<0.474$, and large for $d>0.474$.



\begin{figure}[t]
	\centering
	\includegraphics[width=0.45\textwidth]{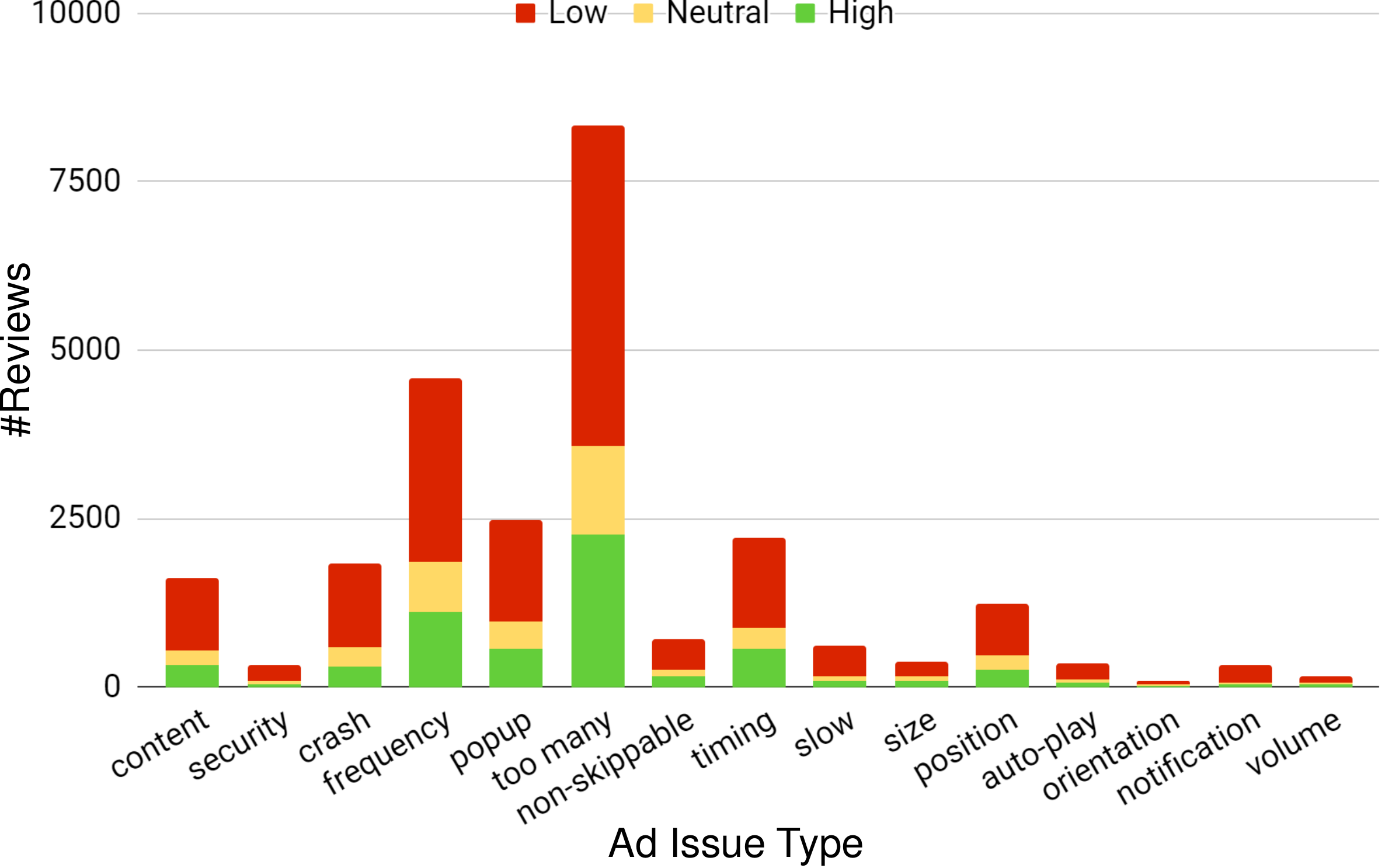}
	\caption{Count distributions of reviews with high, neutral and low ratings across different ad issue types.}
	\label{fig:review_count}
	\vspace{-1em}
\end{figure}

\subsubsection{Findings} 
Figure~\ref{fig:review_count} illustrates the quantity distributions of the 15 ad issues among feedback with high, neutral, and low ratings. Overall, low- and neutral- rated feedback occupies a major proportion (76.11\%) in the ad reviews, implying that \textbf{ad reviews tend to have a negative impact on user ratings}. The result of Chi-Squared test ($\mathit{p-value}=1.94e-44$) indicates that user ratings and ad issue types are strongly correlated, which means that users tend to rate the severity of different issue types differently.

\textbf{The \emph{too many} and \emph{frequency} issue types receive the highest number of negative ratings among all the types.} Focusing on the negative and neutral ad reviews (as shown in Figure~\ref{fig:review_count}), we find that most of them talk about the number of ads (i.e., the \textit{too many} category) and display frequency. These two issue types account for nearly half of all the low-rated reviews (48.37\%). This implies that \textbf{users who are averse to ads mostly complain about the number of ads or their display frequency}. 

\begin{figure}[t]
	\centering
	\includegraphics[width=0.45 \textwidth]{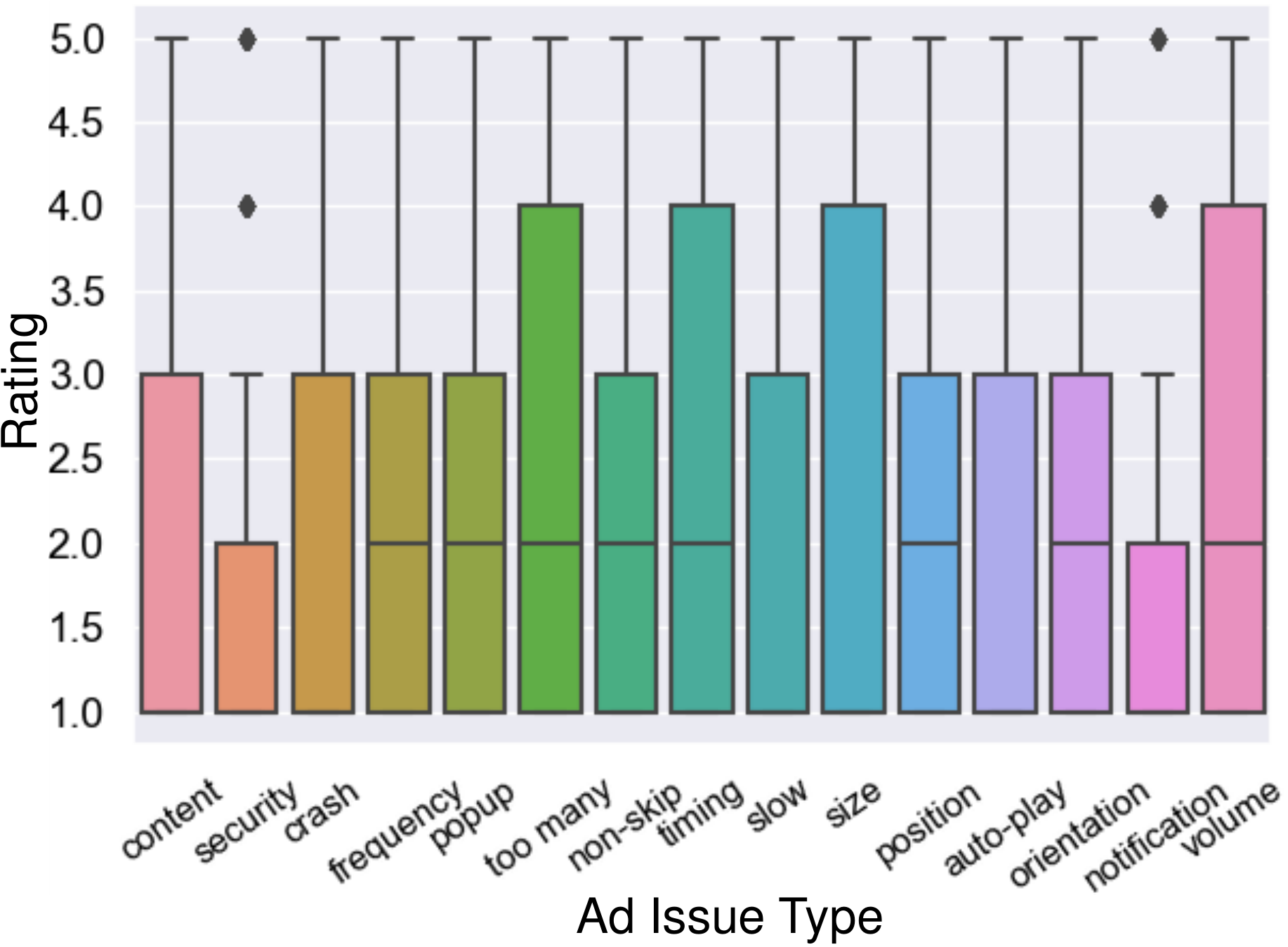}
	\caption{Rating distribution of different ad issue types.}
	\label{fig:rate_topic}
	\vspace{-1em}
\end{figure}

\textbf{Developers should notice the issues related to security and notification.} Figure~\ref{fig:rate_topic} shows the rating distributions of different ad issues. We can observe that most of the ad reviews discussing about specific ad issues are scored with ratings lower than or equal to three, with median star ratings at two. By computing the average scores, we discover that both the \textit{security} and \textit{notification} issues have the lowest ratings (1.8) on average. For example, one one-star-rating review from WeChat says that ``\textit{...Everytime I try to watch or do something, the ad notifications always pop out, and it always directly opens App Store by itself...}''. We further use Mann-Whitney U test to examine whether these two issues receive significantly lower ratings than other issue types respectively. The results of Mann-Whitney U test ($\mathit{p-value}<0.05$) and $d>0.147$ show that both issues have significantly different rating distributions from other issues with at least a small effect size. Thus, developers need to notice the two issue types and try to fix them quickly (more details can be found in Section~\ref{subsec:rq4}).

\textbf{Developer should be cautious about popup and crash-related ad issues.} 
Focusing on the issue types with median values at 1.0 (as shown in Figure~\ref{fig:rate_topic}), we find that the \textit{auto-play}, \textit{popup}, \textit{crash}, \textit{size}, and \textit{slow} issues also correspond to low star ratings besides the \textit{security} and \textit{notification} issues. Considering the percentage distributions obtained in RQ1, we suggest that developers should pay attention to the reviews complaining about \textit{popup} and \textit{crash}, as both constitute of more than 10\% of the ad reviews.


\begin{center}
\setlength{\fboxrule}{1pt}
	\noindent\fbox{
		\parbox{0.46\textwidth}{
			\textbf{Finding 2:}
			Nearly half (48.37\%) of the negative and neutral ad feedback relates to the number of ads and ad display frequency. Besides, developers should pay attention to the ad issues related to popup and crash which tend to receive poorer user ratings and account for more than 10\% of ad reviews. Also, the \textit{security} and \textit{notification}-related ad reviews generally receive lower scores than other types of ad reviews.
			}
		}
\end{center}



\subsection{RQ3: How different are the distributions of ad issue types in different platforms?}
\subsubsection{Motivation}
Popular apps generally publish their products on multiple systems, such as Android, iOS, and Windows. A report in 2018~\cite{crossplatform} showed that the cross-platform app market was expected to hit \$7.5 billion by 2018, and the amount was still on the rise. Users of different platforms may have different preference. 
Also, the two operating systems are different in many aspects, such as Android is more customizable. For maximizing mobile revenue, many popular apps
choose to publish their app versions on multiple platforms, especially App Store and Google Play~\cite{appstores}. Thus, studying the difference of ad issue distributions on the two platforms can help developers weight ad issue types according to the platforms during ad design.

\subsubsection{Methods}
Based on the quantity distributions of ad issue types across platforms per app, we also use Pearson's Chi-Squared test~\cite{pearson1900x} to determine whether two datasets have the same distribution. As a null hypothesis, we make the assumption that
for the same app, the frequency distributions of issue types are similar on different platforms.

\begin{figure}[t]
	\centering
	\includegraphics[width=0.5 \textwidth]{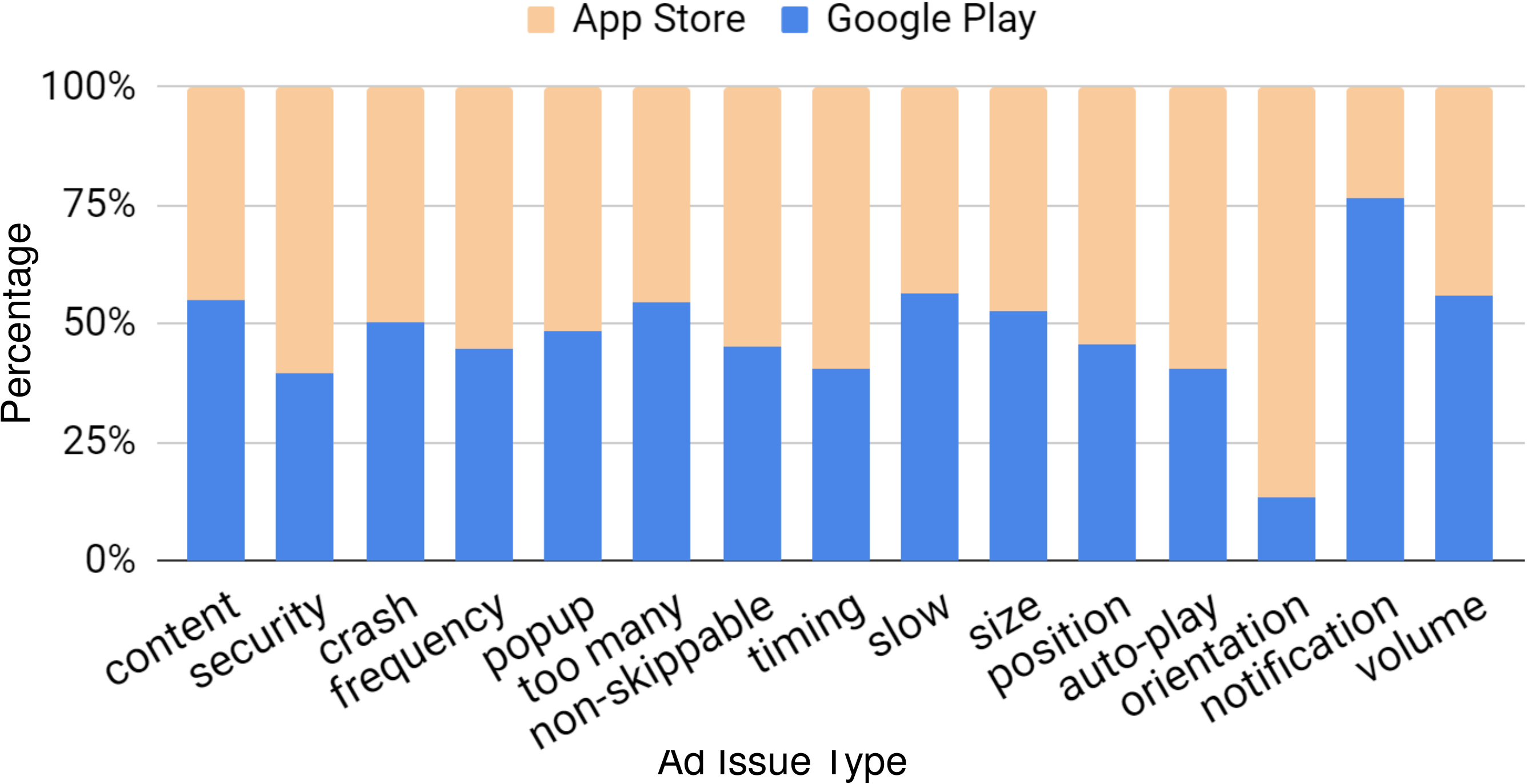}
	\caption{Percentage distribution of ad issue types across App Store and Google Play.}
	\label{fig:platform}
    \vspace{-1em}
\end{figure}

\subsubsection{Findings}

Figure~\ref{fig:platform} shows the percentage distribution of ad issue types on the two platforms. \textbf{We can observe that generally some issues such as \textit{security}, \textit{timing}, \textit{auto-play}, and \textit{orientation} are more complained by iOS users, while other issues including \textit{notification}, \textit{volume}, and \textit{slow} are more concerned by Android users.} By applying Chi-Squared test to the count distributions of ad issue types among the subject apps, we find that all the issue types show significant differences cross platforms, all with $p-value<0.001$. This indicates that the distributions of ad issue types are significantly different on different platforms. \textbf{Developers should design ad maintenance strategies differently for the two platforms.}

\begin{figure}
    \begin{subfigure}[h]{0.48\textwidth}
    \centering
    \includegraphics[width=0.92 \textwidth]{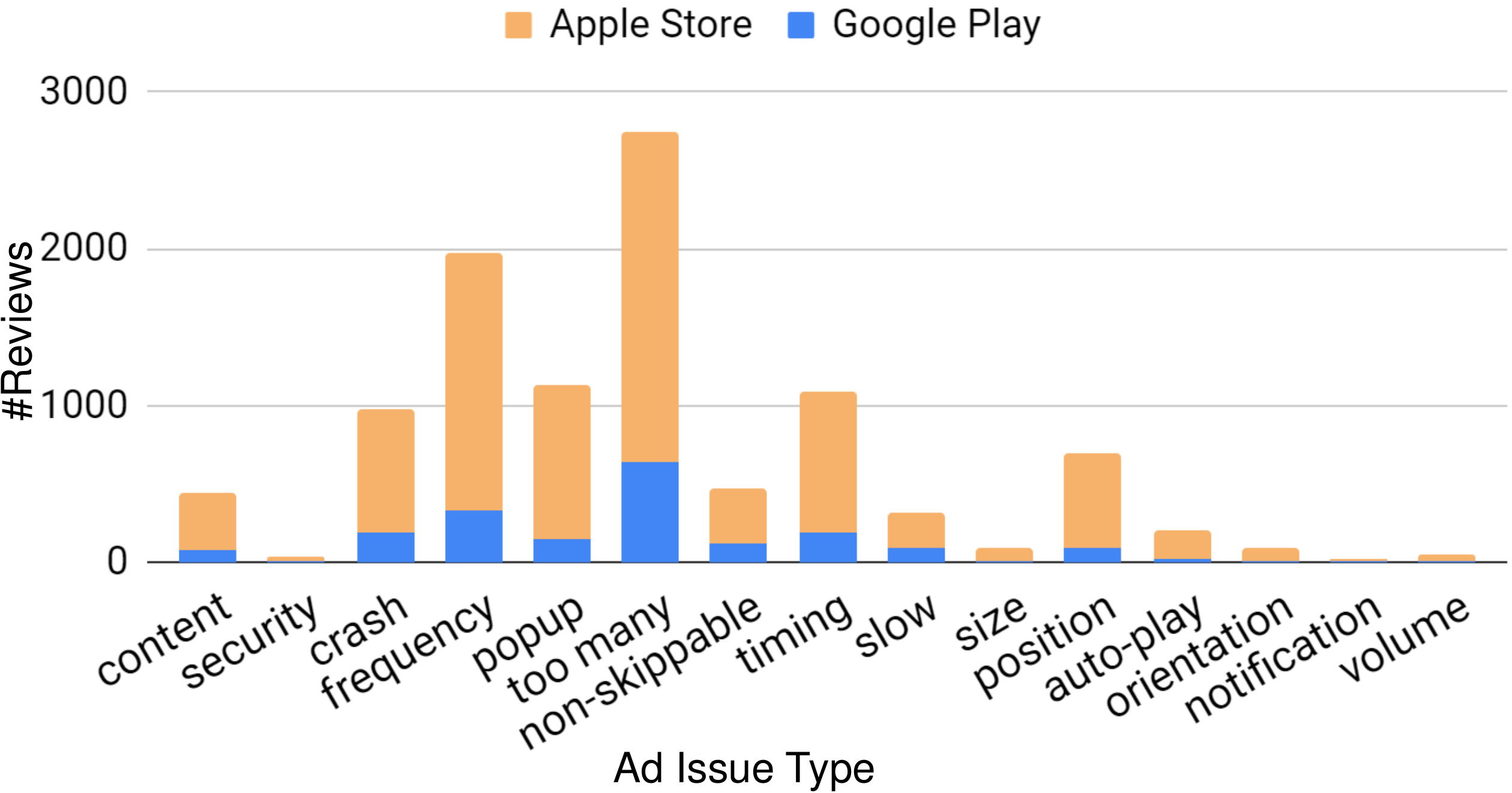}
    \caption{YouTube}
    \end{subfigure}
    \begin{subfigure}[h]{0.48\textwidth}
    \centering
    \includegraphics[width=0.92 \textwidth]{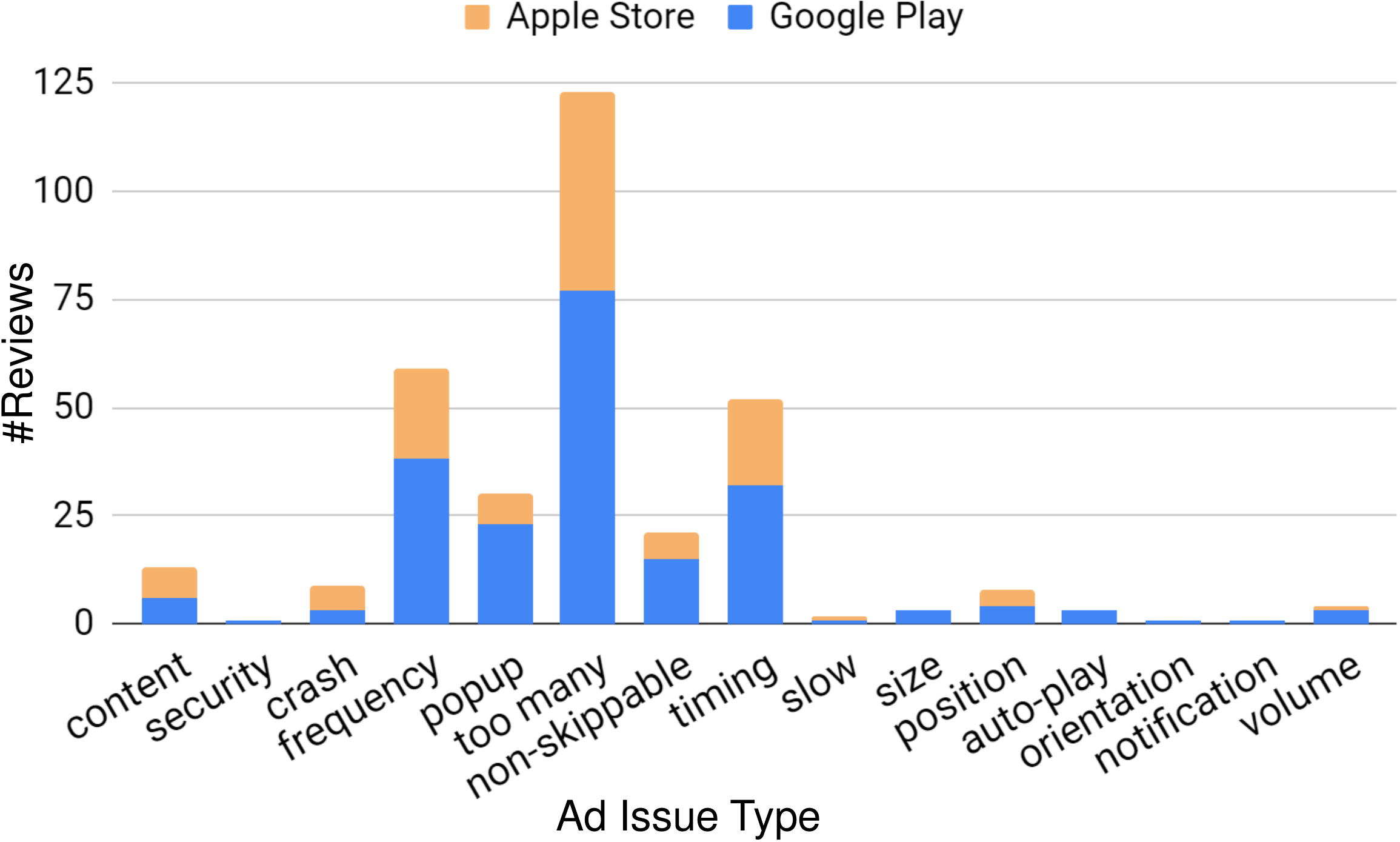}
    \caption{SoundCloud}
    \end{subfigure}
    \caption{Review quantity distributions among ad issue types for YouTube (a) and SoundCloud (b).}
	\label{fig:quantityexample}
	\vspace{-1em}
\end{figure}

We further analyze the quantity distributions of ad issue types for each subject app. 
Figure~\ref{fig:quantityexample} (a) and Figure~\ref{fig:quantityexample} (b) illustrate the review quantity distributions on the issue types for YouTube and Soundcloud, respectively. We can see that the two apps present obviously opposite issue distributions across platforms, e.g., SoundCloud receives more reviews related to the \textit{too many} issue from Google Play than those from App Store, while it is the opposite for YouTube. For SoundCloud, although the difference between issue distributions on both platforms is not statistically significant, its Android app has clearly more ad complaints than its iOS app. This observation further suggests that developers should design ad maintenance strategies according to the deployed platforms.


\begin{center}
\setlength{\fboxrule}{1pt}
	\noindent\fbox{
		\parbox{0.46\textwidth}{
			\textbf{Finding 3:} The quantity distributions of the ad issues show significant differences between App Store and Google Play. For an app, its issue distributions may also behave differently for different platforms.\par
		}}
\end{center}
 
 \subsection{RQ4: What types of ad issues are addressed more quickly by developers?}\label{subsec:rq4}
 \subsubsection{Motivation}
 App developers would address the important app issues feedbacked by users in the updated versions. Similarly, if one ad issue is solved by developers in a timely manner, we can infer that the ad issue is crucial from developers' perspective. We suppose that the developers of popular apps are experienced, and can prioritize issues during maintenance professionally. Thus, the duration of an ad issue can reflect whether the issue type is valued by developers, and provide us additional insights into ad maintenance. 
 
  \begin{figure}[t]
    \begin{subfigure}[h]{0.49\textwidth}
	\centering
    \includegraphics[width=0.9 \textwidth]{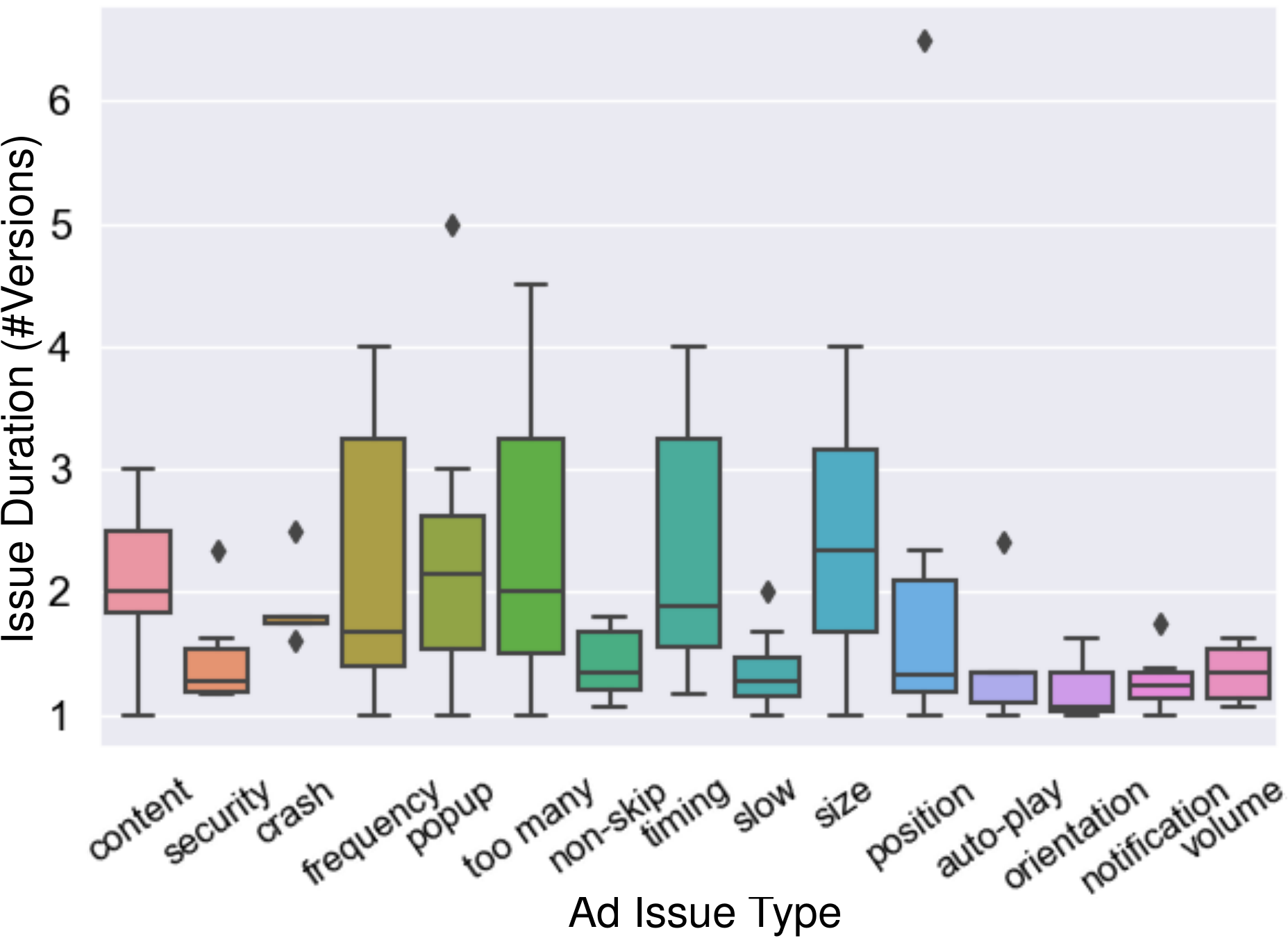}
    \caption{App Store}
    \end{subfigure}
    
    \begin{subfigure}[h]{0.49\textwidth}
    \centering
    \includegraphics[width=0.9 \textwidth]{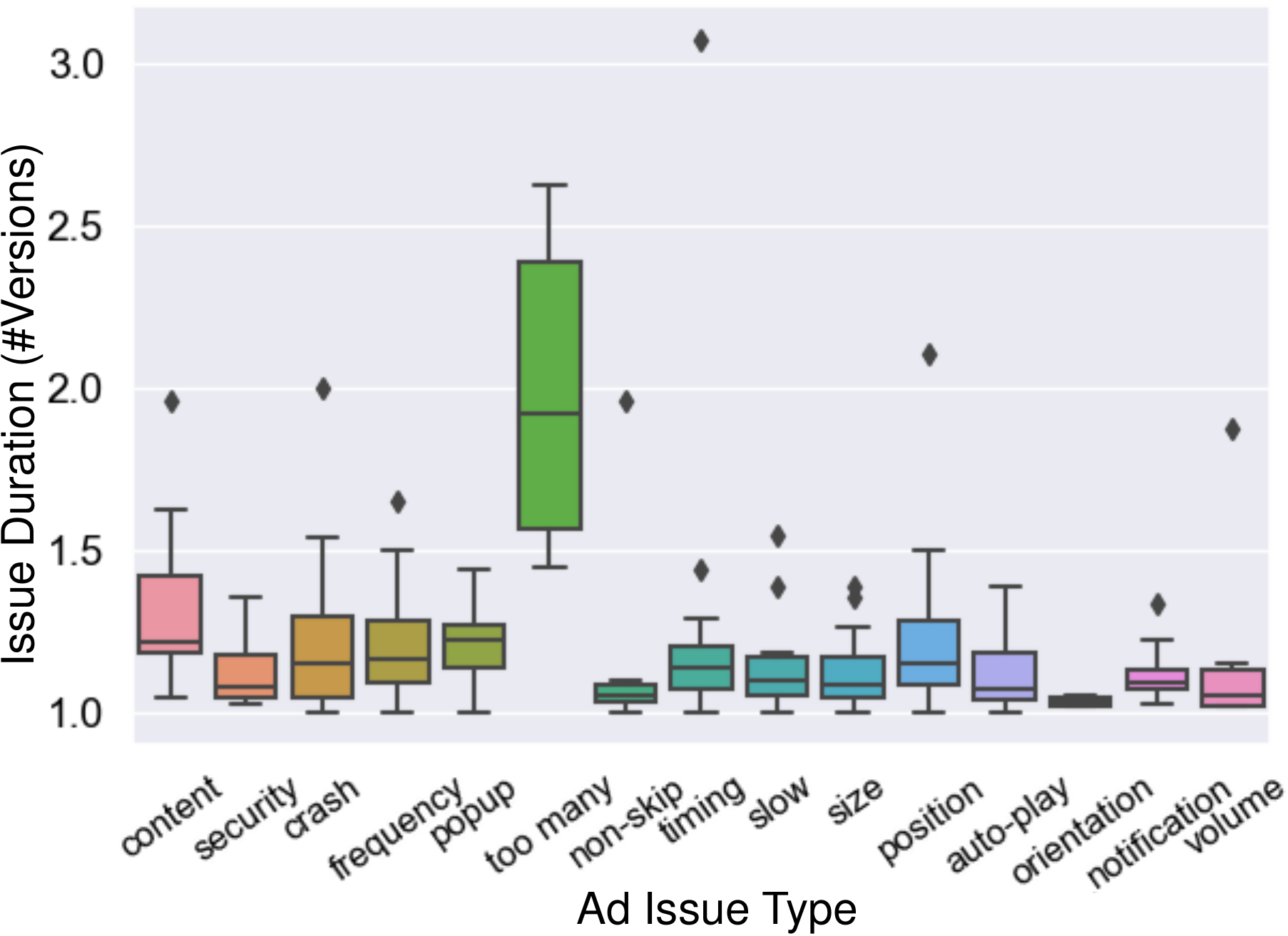}
    \caption{Google Play}
    \end{subfigure}
	\caption{Durations of ad issue types on App Store (a) and Google Play (b). The duration is measured in the number of versions.}
	\label{fig:duration}
\end{figure}

 \subsubsection{Methods}
 We first determine the subject apps for this question by removing those with the number of consecutive versions fewer than three in our collection, since more versions provide us more accurate information about issue changes. As it is challenging to manually check whether an ad issue is indeed fixed in one app version practically, we follow the common strategy used in defect warning analysis~\cite{DBLP:conf/paste/AyewahPMPZ07}.
 In this paper, we define an ad issue is addressed by developers if its percentage is significantly reduced in a version, and not increased in the next version.

We suppose that the percentages of one ad issue over versions $P=\{p_1,...,p_v,...,p_V\}$, where $V$ denotes the total version number, follow a Gaussian distribution $P\sim G(\mu,\sigma)$. An ad issue in one version can be considered addressed if $\frac{p_v-\mu}{\sigma}>\epsilon$, where $\epsilon$ indicates how far the actual value differs from the expected value relative to the typical difference. In statistics, a relative deviation of 2 (i.e., $\epsilon=2$) is often considered as significant~\cite{vu2015mining}. Thus, if $\epsilon>2$ and a decreased trend appears in version $v$ (i.e., $p_v<p_{v-1}$), we can assume that the ad issue is addressed for that version. 
The issue duration $d$ is calculated as the version span between the nearest version $v_n$ with at least one user-review regarding the issue and the current version $v$, i.e., $d=v-v_n$. 
 
 
 
 
 \begin{table}[t]
	\center
	\caption{Number of identified versions for the selected 20 apps.
	}
	\label{tab:version}
	\scalebox{0.85}{\begin{tabular}{l|r|r?l|r|r}
		\hline
		\multirow{3}{*}{\textbf{App Name}} & \multicolumn{2}{c?}{\textbf{Platform}} & \multirow{3}{*}{\textbf{App Name}} & \multicolumn{2}{c}{\textbf{Platform}} \\
		 \cline{2-3} \cline{5-6}
        & \begin{tabular}{@{}c@{}}App \\ Store\end{tabular}  & \begin{tabular}{@{}c@{}}Google \\ Play\end{tabular}  & & \begin{tabular}{@{}c@{}}App \\ Store\end{tabular} & \begin{tabular}{@{}c@{}}Google \\ Play\end{tabular} \\
        \hline
       Candy Crush  &6 & 14& Duolingo &17 & 47\\
       eBay  & 17 & 24 & SHAREit &3 & 50\\
       Amazon  & 5& 17& SoundCloud & 8& 24\\
       Minion Rush  &4 & 5 & Subway Surfers & 7& 20\\
       Netflix  & 9& 55 & TripAdvisor &8 & 11\\
       YouTube  & 14& 99 & Trivia Crack & 20 & 47\\
       My Talking Tom  &3 & 15  & Skype & 9& 40\\
       Twitter  & 35 & 68 & Pinterest  & 12 & 30\\
       Snapchat  &10 & 23 & Viber &5 & 35\\
       Waze  &8 & 20& Instagram & 8& 70 \\
		\hline
	\end{tabular}
}
\end{table}

 \subsubsection{Findings}
 We first remove the apps with fewer versions (i.e., $<3$) or no version information, where the apps will also be removed if we only have their version information on one platform. After this step, we have left 20 apps and 922 versions in total. Table~\ref{tab:version} lists the version number for each app, and Figure~\ref{fig:duration} presents the computed duration distributions among the ad issue types on App Store (upper) and Google Play (below). The result of Mann-Whitney U test~\cite{mann1947test} ($p-value=6.53e-5$) on the average issue durations across platforms shows that developers solve ad issues in significantly different paces for different platforms. As can be observed in Figure~\ref{fig:duration}, issues on Google Play, with average version duration at 1.23 and median duration at 1.19, are generally addressed more quickly than the ones on App Store (avg. 1.78 and med. 1.47).
 
We also find that \textbf{some ad issues 
 would be more quickly addressed by developers than other issue types.} For instance, iOS developers tend to solve \textit{orientation}, \textit{auto-play}, and \textit{notification} issue types more quickly. For Android developers, they would solve the \textit{orientation}, \textit{volume}, and \textit{non-skippable} issue types in the next updated version, with median issue durations at 1.03, 1.05, and 1.05 versions, respectively, as shown in Figure~\ref{fig:duration} (b). 
 Taking an example of YouTube, the app used to receive several user feedback complaining about the non-skippable ads in version 12.01.55. One user commented that ``\textit{I don't have the option to enable non-skippable in-stream ads on my videos, what can I do?}''. 
 The issue also aroused an intense discussion on YouTube Community~\cite{youtubecommunity}, and was finally solved by the developers~\cite{youtubesolved}. In our collected reviews, the number of ad reviews related to the non-skippable issue dropped to zero in the next updated version.

 
 For some ad issues such as \textit{too many} and \textit{content}, both iOS and Android developers may take a longer period to address them. One possible reason is that the ad module is built on specified
 ad provider 
 and the ad content may be difficult for developers to modify. Overall, some ad issue types are more likely to be solved in the next immediate version while other issue types may exist in several consecutive versions.
 

\begin{center}
\setlength{\fboxrule}{1pt}
	\noindent\fbox{
		\parbox{0.46\textwidth}{
			\textbf{Finding 4:} Some ad issue types are prone to be quickly addressed by developers than other ad issues. Also, developers of different platforms for the same app may solve ad issues in different paces.\par
		}}
\end{center}
\section{Implications}\label{sec:lessons}
In this section, we describe the implications of our findings on ad design, ad maintenance, and future research.

\subsection{Implication on In-App Ad Design}

\textbf{Developers should optimize the ad display settings such as the number of ads, display frequency, and display style.} From our analysis, we find that the complaints about the display settings occupy a substantial percentage of studied ad reviews and the the display setting related reviews tend to be accompanied with poor ratings. Developers are suggested to conduct A/B testing to determine an optimal setting for in-app ads. Moreover, strategies such as integrating rewards for watching ads can alleviate users' dislike for ads. It is also worth noting that popup ads appearing during a call or when music is playing, can lead to an unpleasant experience for users, and should be avoided.

\textbf{Developers should carefully design effective strategies to manage ads with long display periods}. We observe that the content and timing-related issues also account for a substantial percentage of ad reviews. Watching long video ads that are not of interest to users would try their patience. Developers should design effective personalization strategy to recommend the right ads of interest to different users. Providing a skip option is another strategy to relieve users' negative emotions.

\subsection{Implication on In-App Ad Maintenance}
\textbf{Developers should ensure app stability as ads are displayed in apps.} Our findings indicate that the crash-related issue appears in a large number of reviews, and corresponds to low user ratings. If the ad libraries are configured incorrectly, the apps' functionalities could be corrupted or slowed down. So we recommend developers to carefully integrate and test the ad libraries before deployment, and to fix the related issue in a timely manner.




\textbf{Developers should prioritize ad issues on different platforms differently.} Our findings demonstrate that the quantity distributions of the ad issue types across different platforms are significantly different. For example, iOS developers tend to solve \textit{orientation}, \textit{auto-play}, and \textit{notification} issue types more quickly than Android developers; while Android developers care more about the \textit{orientation}, \textit{volume}, and \textit{non-skippable} issue types. These results suggest that app developers for a specific platform (Android or iOS) need to put more focus on a subset of ad issues during ad maintenance instead of treating them equally.

\subsection{Implication on Future Research}
\textbf{More empirical research on balancing user experience and ad revenue is needed.} Although anecdotal evidence exists on the adverse impact of in-app ads, unfortunately, few research work has empirically explored how to properly design mobile ads while preserving ad benefits (\textit{e.g.}, click-through rate and ad revenue). We encourage future researchers to perform such studies so that impact of detailed ad design strategies (\textit{e.g.}, choice of ad format and content, ad display frequency, etc.) to ad revenue can be measured and estimated. Developers can then pick ad design strategies in a more informed way by considering the trade-offs of ad revenue and its negative impact to user experience.


\section{Threats to Validity}\label{sec:dis}
\subsection{External Validity}
Threats to external validity concern the possibility to generalize the findings~\cite{noei2019too}. In this work, we consider two platforms, App Store and Google Play, as these two platforms are the two largest global app markets~\cite{appstores}. 
We select 32 apps that exist in the top 100 app charts in both Google Play and App Store as subjects. Hence our results may not generalize to all mobile applications. To mitigate this threat, the apps are selected to cover a broad range of categories and have a significant number of user reviews on both platforms for ensuring their popularity and representativeness. 
Besides, the two platforms may not provide access to all the user reviews. Martin et al.~\cite{MartinMSR15} observed that using incomplete data in app stores may bias the findings. To reduce such a bias on the findings, we collect all the user reviews (i.e., 1,840,349 and 3,243,450 reviews for App Store and Google Play respectively) gradually from September 2014 to March 2019.

\subsection{Internal Validity}\label{subsec:internal}
First, we identify the ad reviews if they contain keywords related to ads, \textit{i.e.}, regex = ad/ads/advert*.
Such strong criterion could lead to significant numbers of true negatives, and might affect the soundness of our findings. To explore the influence caused by the retrieval method, we randomly label a statistically representative sample of 1000 reviews (out of the whole 5,083,799 reviews), providing us with a confidence level of 95\% and a confidence interval of 3\%. The labeling process was conducted by the first author and the second author separately, and reached 100\% agreement rate from both authors. Among the 1000 reviews, five reviews are labeled as related to in-app ads, and our retrieval method can achieve 83.3\% (5/6) and 100\% for precision and recall, respectively. 
This indicates that the regex-based retrieval method can identify ad-related reviews completely.

Second, our manual categorization of the ad reviews is subjected to annotators' bias. We alleviate such threat by following standard card sorting process and making sure that the two annotators agree on the final decision.
Table~\ref{tab:cate_comparison} also compares the ad issue types discovered by Gui et al.~\cite{gui2017arx}, a study on analyzing ad-related complaints, and us. Our detected ad issue types not only cover all the types that are extracted by Gui et al., but also consider a wider range of issues. As shown in Table~\ref{tab:cate_comparison}, Gui et al. did not cover four types. According to our results, not all the four ad issues account for large proportions among the whole ad reviews.

\begin{table}[t]
	\center
	\caption{Difference between the ad issue types that are discovered by Gui et al.~\cite{gui2017arx} and our detected ad issue types.}
	\label{tab:cate_comparison}
	\scalebox{0.88}{\begin{tabular}{l L{5cm}}
		\hline
		\begin{tabular}{@{}l@{}}\textbf{Ad Issue Types} \\ \textbf{by Gui et al.~\cite{gui2017arx}}\end{tabular} & \multicolumn{1}{c}{\textbf{Our Ad Issue Types}} \\
		\hline
		\multicolumn{2}{c}{\textbf{The Ad Issue Types in Common}} \\\hline
		Frequency & Frequency \\
		Content & Content \\
		Popup & Popup \\
		Timing & Timing \\
		Size & Size \\
		Location & Position \\
		Crash & Crash \\
		Slow & Slow \\
		Notification & Notification \\
		Auto & Merged to our \textit{Auto Play} issue type \\
		Video & Removed as unspecific ad issues \\
		Intrusive & Removed as unspecific ad issues \\
		Block & Merged to our \textit{too many} issue type \\
		Privacy & Merged to our \textit{security} issue type \\
		Paid & Removed as unspecific ad issues \\
		Battery & Merged to our \textit{slow} issue type due to its tiny percentage \\
		 \hline
		\multicolumn{2}{c}{\textbf{The Ad Issue Types not in Common}} \\\hline
		Not Covered & Too Many \\
		Not Covered & Non-skippable \\
		Not Covered & Volume \\
		Not Covered & Orientation \\
		\hline
	\end{tabular}
}
\end{table}

Third, our effort to automatically categorize numbers of ad reviews could potentially raise some questions. Especially, the categorization can be tainted by the initial choice of keywords. Also, users express the same issues in various ways. To mitigate the threat, we evaluate our classification against the manual annotation of 280 ad reviews, as discussed in Section~\ref{subsec:auto_class}. Regarding the classification methods, we compare the adopted algorithm, \textit{i.e.}, combining Classifier Chains approach with Support Vector Machine (CC+SVM), with other typical multi-label classifiers, including random weighted classifier~\cite{DBLP:journals/ese/BaoXXLH18}, $K$ nearest neighbors (KNN) algorithm\footnote{we set $K=5$ via cross-validation.}~\cite{DBLP:conf/nips/WeinbergerBS05}, and also CC jointly trained with Logistic Regression (CC+LR)~\cite{DBLP:books/wi/HosmerL00}. Table~\ref{tab:comparison} presents the comparison results. The results demonstrate that Classifier Chains (CC) algorithms have a better performance when compared to Random Weight and KNN algorithms. For the basic estimator of CC, SVM shows a better performance than LR. Therefore we choose CC with SVM as our multi-label classifier in our study.


\begin{table}[t]
	\center
	\caption{Comparison results of different multi-label classifiers.}
	\label{tab:comparison}
	\scalebox{0.88}{\begin{tabular}{l|r|r}
		\hline
		    Method & Precision & Recall \\
		\hline
		    Random Weight & 78.43\% & 75.18\% \\
		    KNN & 68.92\% & 66.01\% \\
		    CC+LR & 87.13\% & 79.54\% \\
		    CC+SVM & \textbf{91.18\%}& \textbf{85.57\%} \\
		\hline
	\end{tabular}
}
\end{table}

\subsection{Construct Validity}
There are several approaches to understand what aspects users are complaining about mobile in-app ads. For example, interviewing and surveying mobile users might be one way. In this paper, we chose to instead look at the actual user feedback. Both approaches have their benefits and limitations. For example, with surveys, users might miss reporting on some ad issue types since we are depending on their collection. Nevertheless, a mining approach might be limited since the collected data cannot represent all issue types related to ads. Thus, we suggest that future studies are needed to triangulate our findings through user surveys.
\section{Related Work}\label{src:literature}

\subsection{App review analysis}
App review analysis explores the rich interplay between app customers and their developers \cite{FINKELSTEIN2017119}. The analysis has been proven helpful and significant in various aspects of app development.

Iacob et al.~\cite{DBLP:conf/bcshci/IacobVH13} manually label 3,278 reviews of 161 apps, and discover the most recurring issues users report through reviews. Since mining app reviews manually is labor-intensive due to the large volume, more attempts on automatically extracting app features are conducted in prior studies. For example, Iacob and Harrison \cite{DBLP:conf/msr/IacobH13} design MARA for retrieving app feature requests based on linguistic rules. Man et al. ~\cite{DBLP:conf/issre/ManGLJ16} propose a word2vec-based approach for collecting descriptive words for specific features, where word2vec~\cite{DBLP:conf/nips/MikolovSCCD13} is utilized to compute semantic similarity between two words. Vu et al.~\cite{vu2015mining,vu2016phrase} have investigated how to facilitate keyword retrieval and anomaly keyword identification by clustering semantically similar words or phrases. Another line of work focuses on condensing feature information from reviews and captures user needs to assist developers in performing app maintenance~\cite{di2016would,DBLP:conf/icse/VillarroelBROP16}. Maalej and Nabil~\cite{DBLP:conf/re/MaalejN15} adopt probabilistic techniques to classify reviews into four types such as bug reports and feature requests. Di Sorbo et al.~\cite{di2016would} build a two-dimension classifier to summarize user intentions and topics delivering in app reviews. \cite{guzman2014users}, \cite{DBLP:conf/kbse/GuK15}, and \cite{DBLP:conf/www/LuizVAMSCGR18} concentrate on specific app features and propose methods to identify corresponding user sentiment or opinions. There are also review-based explorations aiming at supporting the evolution of mobile apps ~\cite{paid2015cygao,cuiyun2018idea,DBLP:journals/jss/PalombaVBOPPL18}. Specifically, Palomba et al.~\cite{DBLP:journals/jss/PalombaVBOPPL18} trace informative crowd reviews onto source code changes to monitor what developers accommodate crowd requests and users' follow-up reactions as reflected in user ratings. They observe that developers implementing user reviews are rewarded in terms of significantly increased user ratings. Other research considers device- or platform-specific app issues~\cite{DBLP:journals/ese/McIlroyAKH16,DBLP:conf/icse/LuLLXMH0F16,DBLP:conf/issre/ManGLJ16}. We refer to Martin \textit{et al.}'s survey for an extensive overview of mobile app store analysis research~\cite{DBLP:journals/tse/MartinSJZH17}.



\subsection{User perceptions of in-app ads}

According to the research \cite{DBLP:journals/software/KhalidSNH15}, privacy \& ethics and hidden cost are the two most negatively perceived complaints (and are mostly in one-star reviews) among all studied complaint types. An interesting empirical study by Gui \textit{et al.}~\cite{gui2015truth} exhibits obvious hidden costs caused by ads from both developers' perspective (\textit{i.e.}, app release frequencies) and users' perspective (\textit{e.g.}, user ratings). \cite{wei2012profiledroid} and \cite{nath2015madscope} discover that the ``free'' nature of apps comes with a noticeable cost by monitoring the traffic usage and system calls related to mobile ads. Ullah~\textit{et al.}~\cite{DBLP:conf/infocom/UllahBKK14} also find that although user's information is collected, the subsequent usage of such information for ads is still low.
To alleviate these threats, \cite{mohan2013prefetching} and \cite{vallina2012breaking} develop a system to enable energy-efficient ad delivery. Gui \textit{et al.}~\cite{DBLP:conf/greens/GuiLWH16} propose several lightweight statistical approaches for measuring and predicting ad related energy consumption, without requiring expensive infrastructure or developer effort. Gao \textit{et al.}~\cite{GaoAmobile18} investigates the performance costs raised by different advertisement schemes, and demonstrates that some ad schemes that produce less performance cost and provide suggestions to developers on ad scheme design. Ruiz \textit{et al.}~\cite{ruiz2014impact} also find that integrating certain ad libraries can negatively impact an app's rating. In Gui \textit{et al.}~\cite{gui2017arx}'s work, ad-related complaints are extracted from manually annotating 400 user reviews. Different from the prior work, we focus on analyzing ad issues based on a large-scale user review corpus and considering multiple factors such as fix durations and app platforms.


\section{Conclusion}\label{sec:conclusion}
Inappropriate ad design could adversely impact app reliability and ad revenue. Understanding common in-app advertising issues
can provide developers practical guidance on ad incorporation.

In this paper, we have presented a large-scale analysis on ad reviews to summarize common issues of in-app advertising. We discover the common ad issue types by manual annotation. Based on the automatic categorization results of a large-scale ad reviews, we observe the general distributions of the ad issue types, the relations between ad issue types and user ratings, the distributions of ad issues across platforms, and fix durations. We summarize our findings and their implications to app developers for more effective and reliable design and maintenance of in-app ads. In the future, we will consider other aspects such as mobile device types to gain further insights about the impact of in-app advertising on app reliability.



\ifCLASSOPTIONcaptionsoff
  \newpage
\fi



\bibliographystyle{IEEEtran}
\bibliography{main}
%



%





\end{document}